# Machine Learning and Finite Element Method for Physical Systems Modeling

O. Kononenko, I. Kononenko
oleksiy.s.kononenko@gmail.com

*Modeling of physical systems includes extensive use of software packages that implement the accurate finite element method for solving differential equations considered along with the appropriate initial and boundary conditions. When the problem size becomes large, time needed to solve the resulting linear systems may range from hours to weeks, and if the input parameters need to be adjusted, even slightly, the simulations have to be re-done from scratch. Recent advances in machine learning algorithms and their successful applications in various fields demonstrate that, if properly chosen and trained, these models can significantly improve conventional techniques. In this note we discuss possibilities to complement the finite element studies with machine learning and provide several basic examples.*

**Motivation**

Nowadays, the vast majority of analysis in structural mechanics, fluid dynamics, electromagnetics and many other areas is based on the finite element method (FEM) for solving boundary value problems. The approximate solution of the corresponding partial differential equation can be computed at the discrete number of points over the computational domain through the analysis of the resulting linear algebraic system. In some cases for the time domain problems, the resulting linear system must be solved at each time step.

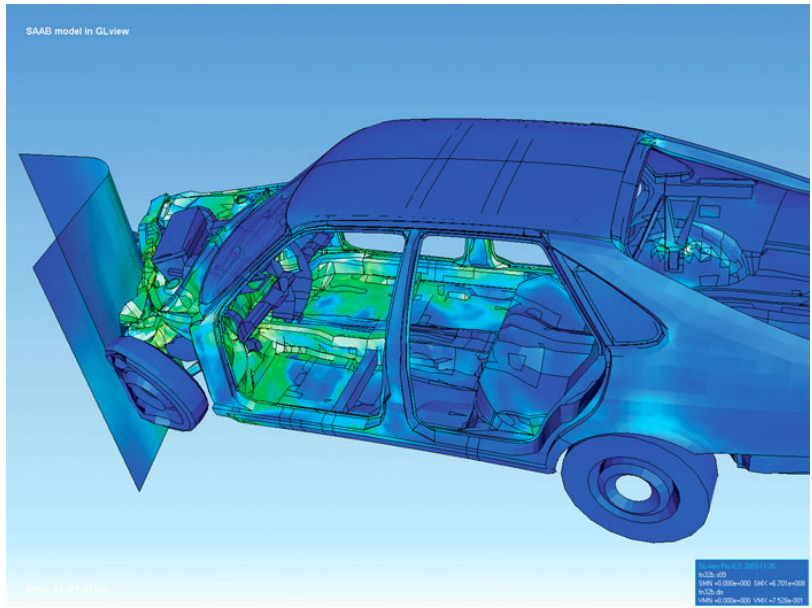

Figure 1. Visualization of the car deformations in an asymmetrical crash based on the finite element analysis [1].

For the realistic applications, see Fig. 1, the size of the system can be extremely large, i.e. from millions to billions degrees of freedom, and the simulation time on a cluster or supercomputer can vary from hours to days or weeks. No matter what is the scalability of the linear solver, the finite element modeling requires massive computational resources and, except for the final results, the machine experience gained during the simulation is lost. It means that when the input has to be adjusted, even slightly, or one needs to re-do the study done elsewhere, in most of the cases the time-consuming analysis has to be done from scratch.

From the other hand, if the physical system is properly discretized, the finite element results are very accurate and, along with the input parameters, can be used to train the machine learning model. One of the options to do this efficiently is to train the model on the large sets of data generated by the well-developed conventional FEM tools on the random basic problems, see Fig. 2. The training data can also be complemented by the actual measurements as well as by the simulation results for the realistic problems shared by the users of the FEM packages. When properly trained, this model can be used for a wide range of other applications.

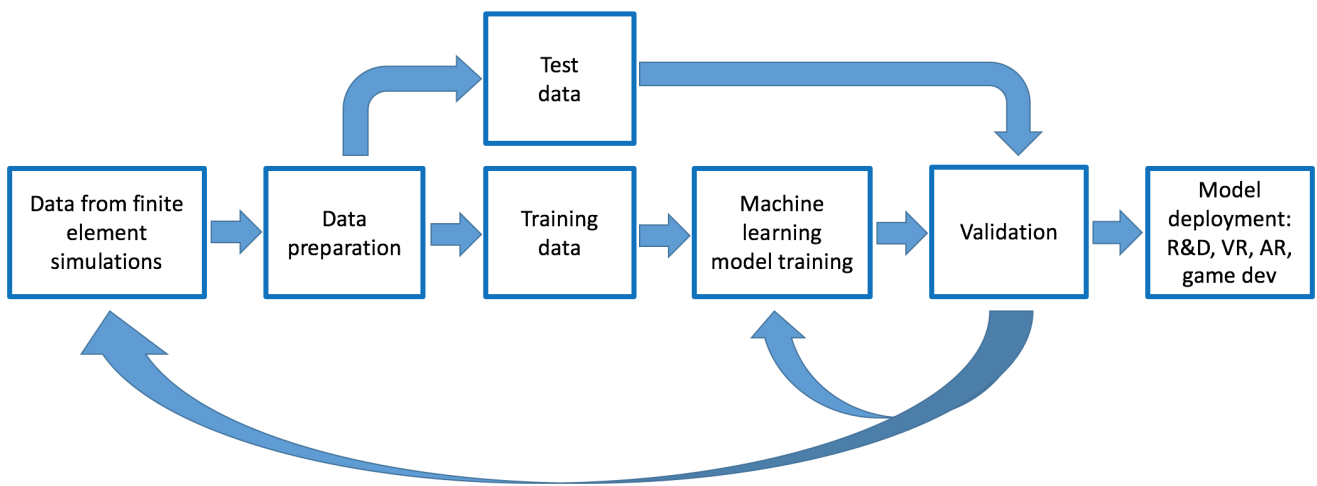

Figure 2. A proposed workflow to train the machine learning model based on the data generated by the finite element software on the random basic problems.

For this purpose utilization of deep neural networks may be advantageous for several major reasons: many physical systems are highly non-linear and require uncertainty quantification. In contrast to the other machine learning applications, where clean data and automated supervision may be an issue, the finite element based approach is pretty unique. Moreover, if problems from different fields are governed by the same type of equations, that is common for electromagnetics and structural mechanics, for example, the learning process can be greatly simplified. Even though some efforts have been already made in this direction [2-5], as for today, all these studies are kind of fragmentary and there is no publicly available machine learning tool capable to replace or outperform the finite element simulators.

**Examples**

In this note we illustrate this approach by considering several basic examples. First, the one-dimensional harmonic oscillator, one of the most fundamental systems that, when displaced from its equilibrium, experiences a restoring force proportional to the displacement. We have trained the artificial neural network on the data generated from the analytical solution of the harmonic oscillator equation, and use it to predict the amplitude of the oscillator as a function of the driving frequency, see Example 1.

In Example 2 we study a response of the three-dimensional beam to the harmonically varying load. In this case the 3D finite element analysis is done for a range of frequencies and the network is trained on the generated multidimensional data to predict the maximum displacements of the system in all three directions.

***Example 1. One-dimensional harmonic oscillator.***
In this example the amplitude of a generic 1D harmonic oscillator is computed analytically for a set of driving angular frequencies up to 10 Hz. The neural network consisting of two hidden layers is trained on a random subset and used to predict the response of the oscillator to harmonic loading at particular frequency, see Fig. 3.

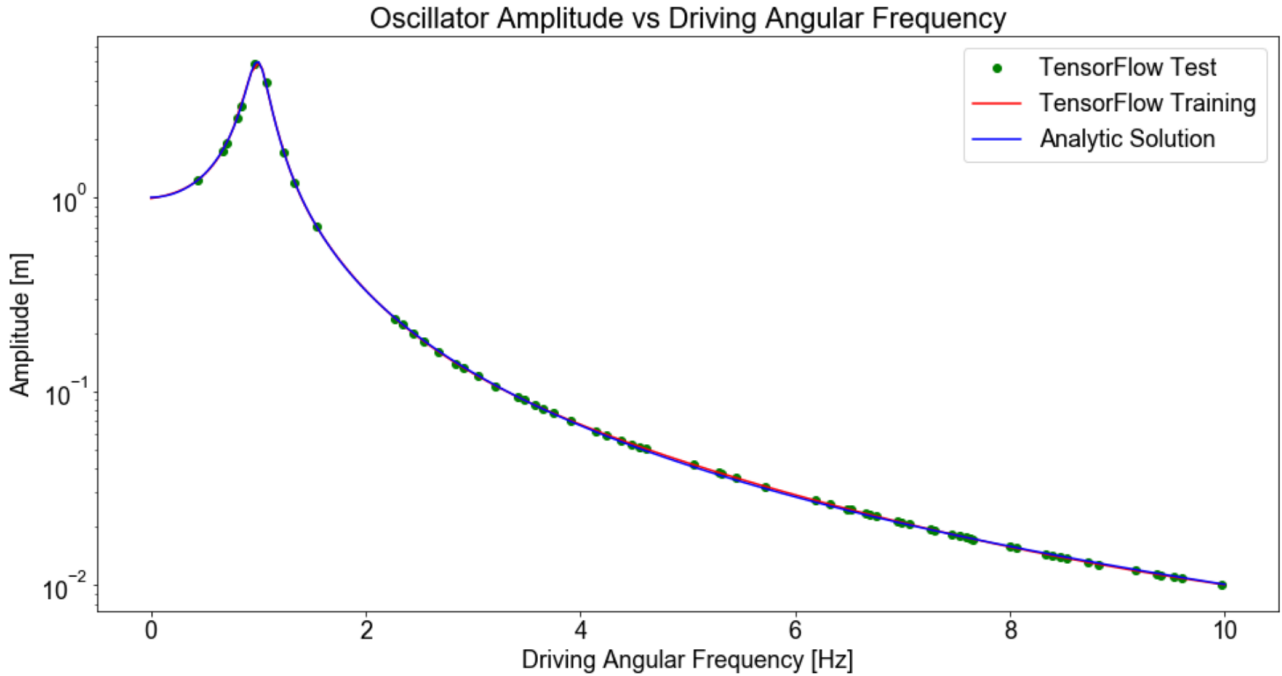

Figure 3. A comparison of the harmonic oscillator amplitude computed with the neural network on the test and training data against the analytic solution.

If the network is properly designed and trained a very good agreement between the analytical solution and the model prediction can be reached. In this example the network had one input (frequency) and one output (amplitude), as well as two hidden layers both with the height of 100 neurons. The stochastic gradient descent method, as implemented in TensorFlow, was used for the minimization of the mean squared error, see [6] for details.

***Example 2. Three-dimensional beam under a harmonically varying load.***
As a little bit more advanced example, we consider a 3D stainless steel beam and calculate its response to a harmonically varying external loading. First, the associated analysis is done by using the finite element code for the range of frequencies up to 200 Hz, and then the artificial neural network was trained on the random frequency subset to predict the maximum displacement of the system in all three dimensions: *x*, *y* and *z*, see Fig. 4. In contrast to Example 1, in this case the model was trained to take into account multiple resonant modes (peaks) in different directions, see Fig. 5.

The designed network had one input (frequency), two hidden layers both with the height of 200 neurons and three outputs (maximum displacements in *x*, *y* and *z*). Due to the features having a broad range of values, a logarithmic scaling was applied. The Adam optimization algorithm, extension to stochastic gradient descent implemented in TensorFlow, was used for the minimization of the mean squared error, see [6].

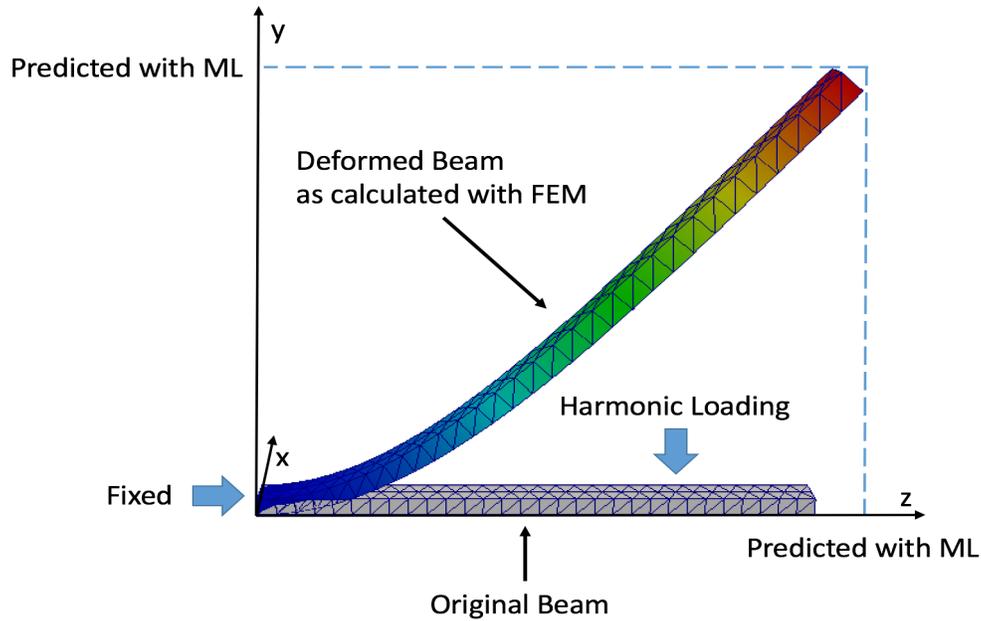

Figure 4. The original geometry of the stainless steel beam featuring the finite element mesh and the boundary conditions; the deformed geometry and the machine learning predictions of the maximum displacements at the driving frequency of 9 Hz.

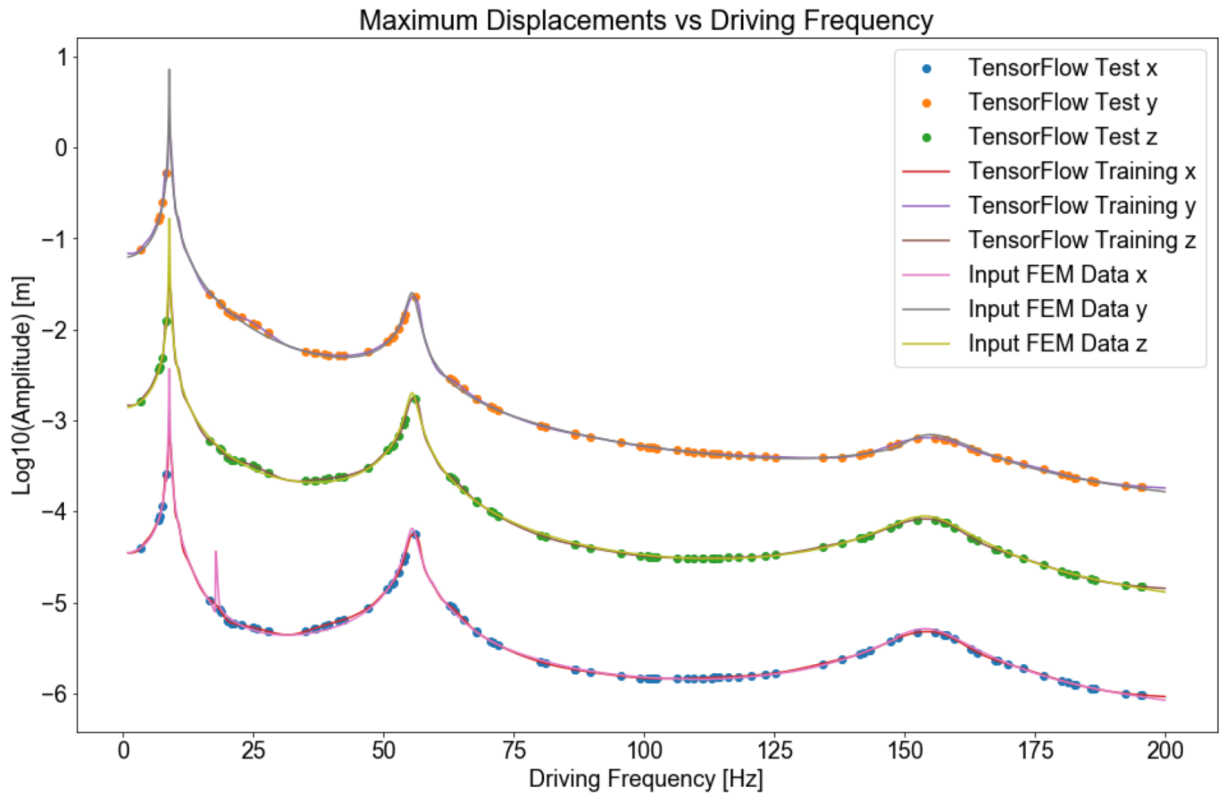

Figure 5. A comparison of the maximum beam displacements in *x*, *y* and *z* as computed with the artificial neural network on the test and training data against the original input from finite element simulator.

## Conclusions

Obviously, real applications will be much more complicated than the examples demonstrated, and will require training on the extremely large datasets generated by the FEM software for various problems, including different combinations of the boundary and initial conditions;  not just the maximum displacements, but, for example, the full field of deformations or electromagnetic fields on all the nodes of the computational mesh.

Since the ultra-high dimensional output is required, the neural network design is a real challenge, that can be facilitated by involving domain experts in the machine learning research. However, without any doubts, this approach can be advanced to solve more realistic problems [7] in time and frequency domains, and, when this kind of the machine learning model is built, it will have an enormous number of applications in sciences, virtual and augmented realities, game development, and many other areas.

## References

[1] Impact Finite Element Program Suite, http://www.impact-fem.org/
[2] T. Nguyen-Thien, T. Tran-Cong, Approximation of functions and their derivatives: A neural network implementation with applications, In Applied Mathematical Modelling, Volume 23, Issue 9, 1999, Pages 687-704.
http://www.sciencedirect.com/science/article/pii/S0307904X99000062?via%3Dihub
[3] Javadi, A.A., Tan, T.P. and Zhang, M.X. (2003), "Neural network for constitutive modelling in finite element analysis", Computer Assisted Mechanics and Engineering Sciences, Vol. 10, pp. 523-9.
[4] J. D. Martín-Guerrero et al., Machine Learning for Modeling the Biomechanical Behavior of Human Soft Tissue, 2016 IEEE 16th International Conference on Data Mining Workshops (ICDMW), Barcelona, 2016, pp. 247-253. http://ieeexplore.ieee.org/document/7836673/
[5] L. Ladicky et al., Physicsforests: real-time fluid simulation using machine learning. In ACM SIGGRAPH 2017 Real Time Live! (SIGGRAPH '17). ACM, New York, NY, USA, 22-22. DOI: https://doi.org/10.1145/3098333.3098337
[6] https://github.com/oleksiyskononenko/mlfem/blob/master/ML_FEM.ipynb
[7] Physics Forests, Real-time Waterfall, https://www.youtube.com/watch?v=jjYHqakTrWY